\documentclass[preprint]{elsarticle}

\usepackage{lineno,hyperref}

\begin{document}

\begin{frontmatter}

\title{Double-Fano resonance  in a two-level quantum system coupled to zigzag Phosphorene nanoribbon}



\author[mymainaddress]{Mohsen Amini\corref{mycorrespondingauthor}}
\cortext[mycorrespondingauthor]{Corresponding author}
\ead{msn.amini@sci.ui.ac.ir} 

\author[mymainaddress]{Morteza Soltani}
\author[mysecondaryaddress]{Samira Baninajarian}
\author[mymainaddress]{Mohsen Rezaei}


\address[mymainaddress]{Department of Physics, University of Isfahan (UI), Isfahan 81746-73441, Iran}
\address[mysecondaryaddress]{Faculty of Physics, University of Kashan, Kashan 87317-53153, Isfahan, Iran.}

\begin{abstract}
Double-level quantum systems are good candidates for revealing coherent quantum transport properties.
Here, we consider quantum interference effects due to the formation of a  two-level system (TLS) coupled to
the edge channel of a zigzag Phosphorene nanoribbon (ZPNR).
Using the tight-binding approach, we first demonstrate the formation of a TLS in bulk Phosphorene sheet due to the existence of 
two nearby vacancy impurities.
Then, we show that such a TLS can couple to the quasi-one-dimensional continuum of the edge states in a ZPNR which results in the 
the appearance of two-dip Fano-type line shapes.  
To this end, we generalize the Lippmann-Schwinger approach to study the scattering of edge electrons in a ZPNR  by two coupled impurity defects.   
We obtain an analytical expression of the transmission coefficient which shows that the positions and widths of the anti-resonances can be controlled by changing the intervacancy distance as well as their distance from the edge of the ribbon.
This work constitutes a clear example of the multiple Fano resonances in mesoscopic transport.
\end{abstract}


\end{frontmatter}


\section{Introduction}
Fano resonance is an important type of resonant scattering phenomena which originates from the interference of a discrete state and a continuum of states~\cite{Fano}. The main feature of Fano resonances is the appearance of an asymmetric resonance profile in the transmission spectrum which is distinct from that of symmetric Lorentz line shapes.
The Fano effect, as a universal phenomenon, has been reported in different areas of physics, 
like atomic and photonic physics~\cite{Waligorski, Limonov}, condensed matter systems~\cite{Bulka}, 
metamaterials and nonlinear optics~\cite{Lukyanchuk, Samson}, and so on.
One of the most relevant systems to observe Fano resonances are the low-dimensional electronic quantum devices in which quantum dots and quantum wells are coupled to semiconductor nanostructures.   
In particular, as an important case, quantum dots side-coupled to quantum wires are studied for different coupling manners both experimentally and theoretically~\cite{Kobayashi-2004, Kobayashi-2002, Kobayashi-2005, Johnson, Orellana, Ricco, Yang, Ladron}.
The appearance of Fano resonances in quantum dot structures is a result of the existence of bound states embedded in the continuum
which was earlier proposed by von Neumann and Wigner~\cite{Wigner}. 
Since then, a large number of studies have explored the possibility of the formation of the bound states in the continuum in different setups.


On the other hand, the possibility of controlling Fano lineshapes by impurities or structural defects is addressed in~\cite{Tanaka, Garmon,Reyes}. 
Apart from these one-dimensional quantum wires, quasi-one-dimensional lattices with peculiar electronic states 
around the Fermi level have become relevant~\cite{Bahamon}.
In this regard, ZPNRs are promising candidate materials because of the presence of conducting channels at the edges of the ribbon. 
This is due to the formation of quasi-flat edge bands in ZPNRs which are entirely isolated from the bulk bands~\cite{Ezawa}.
For this system, it is known that the presence of point defects including vacancy and impurity changes the electronic transmission through such
edge channels~\cite{Asgari,Peeters,Amini1}.
Interestingly, in the presence of a single vacancy defect, electronic transmission through the edge channels of ZPNRs shows Fano antiresonances which can be tuned by changing the distance of the vacancy with respect to the edges~\cite{Amini2}.
Therefore, It would be of great interest to further explore such an effect in the generation of multiple Fano resonances due to the
existence of multiple discrete states.


In the present work, we aim to study theoretically the
the problem of vacancy-driven multiple Fano resonances in a ZPNR.
We consider the scattering of edge states of a ZPNR by the presence of two nearby vacancies located close to the
the boundary of the ribbon.
To do so, we use the tight-binding description of the electronic states in a ZPNR~\cite{Ezawa} which is developed to obtain the localized states associated with a single vacancy defect in a ZPNR~\cite{Amini1} (by two authors of the present study). We first generalize this approach to the case of states localized at two separate yet nearby vacancies. 
We observe that the two nearby vacancies could be explained as a TLS (or equivalently a double-level quantum dot) for which the corresponding localized states form a symmetric and an antisymmetric bound state within the ZPNR edge spectrum. 
Then, we employ the Lippmann–Schwinger equation to study the scattering of the conducting edge electrons from such TLS  when the vacancies are located close to the edge of the ribbon.
We show that the resulting double-dip Fano anti-resonances in the transmission of the edge channels of ZPNRs is connected to the formation of
discrete bound states induced by two vacancy defects close to the boundaries of ribbons.
Next, we explain this result by comparing the results obtained for transmission through a double-impurity system side-coupled to a quantum wire.

The rest of the paper is organized as follows. In Sec.~\ref{SecII}  we obtain the analytical formulas characterizing the wave function associated with two nearby vacancy defects in a ZPNR.
We also discuss the possible realization of a TLS formed by coupling between such vacancy defects in a ZPNR.
In Sec.~\ref{SecIII} we generalize the Lippmann–Schwinger approach to study the scattering of conducting electrons from two localized impurities analytically. 
Based on these analytical expressions, we calculate and discuss the transmission coefficient of a ZPNR with two vacancy defects near the edge of the ribbon which shows Fano lineshapes with two minima.  
Finally, the main results are
summarized in Sec.~\ref{SecIV}.

\section{Localized states due to the presence of two nearby vacancy defects}
\label{SecII}

\begin{figure}[t!]
  \centering
  \includegraphics[width=8cm]{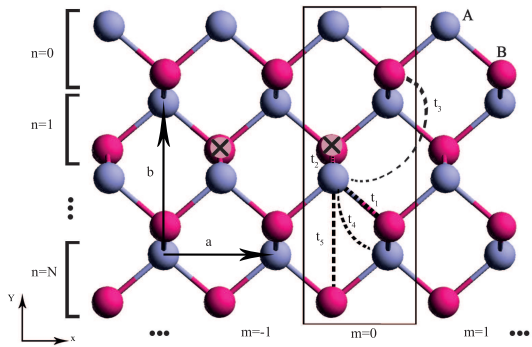}
  \caption{Realspace representation of the lattice structure of Phosphorene. 
The lattice constant $a (b)$ is considered in the $x (y)$ direction.  
$m$ and $n$ are the indexes of the vertical armchair chain along the
horizontal direction, and the horizontal zigzag chain along
the vertical direction, respectively. Each
lattice site can be described by $(n, m, \nu)$ with $\nu = A, B$ referring to the sub-lattices.
Also, each symbol $\times$ shows a lattice site in which there is a single vacancy impurity.
}
\label{F1}
\end{figure}

In this section, we consider the non-planar puckered structure
of phosphorene (see Fig.~\ref{F1} ) 
which can be well described by the following tight-binding Hamiltonian~\cite{Katsnelson}
\begin{equation}\label{E1}
\hat{H}_{Ph} = \sum_{\langle i,j \rangle} t_{ij} c^{\dag}_{i} c_{j} + h.c.\;\;.  
\end{equation}
The summation runs over all lattice sites of the system and $\langle i,j \rangle$ represents summation over only considered neighbors. 
$t_{ij}$ represents the hopping energy between sites $\textit{i}$ and $\textit{j}$ and  $c_{i}^{\dag} (c_{i})$ is the electron creation (annihilation) operator in site $\textit{i}$, and \textit{h.c.}  stands for Hermitian conjugate. 
While it is shown that considering the five most important hopping integrals $ t_{1}=-1.220eV $, $ t_{2}=3.665eV $, $ t_{3}=-0.205eV $, $ t_{4}=-0.105eV $, $ t_{5}=-0.055eV $ is sufficient to provide a reasonable description of the phosphorene band structure, it is more convinient and reasonable to keep only three hopping parameters $t_1, t_2$ and $t_4$ for analytical calculations~\cite{Ezawa,Amini1}.
Therefore, we can rewrite the Hamiltonian of Eq.~({\ref{E1}}) keeping only the important terms of first, second and fourth nearest neighbors as
\begin{equation}\label{E2}
\hat{H}_{Ph}=\hat{H}_{0}+\hat{H}_{1}  ,
\end{equation}
\begin{equation}\label{E}
\hat{H}_{0}=t_{1}\sum_{\langle i,j\rangle_{1st}}c_{i}^{\dag}c_{j}+t_{2}\sum_{\langle i,j\rangle_{2nd}}c_{i}^{\dag}c_{j}+h.c.,  
\end{equation}
\begin{equation}\label{E4}
\hat{H}_{1}=t_{4}\sum_{\langle i,j\rangle_{4th}}c_{i}^{\dag}c_{j}+h.c. \;\;. 
\end{equation}

It is known that the Phosphorene sheet can be terminated along with the $y$-direction in Fig.~\ref{F1} to form a zigzag nanoribbon. 
Using a plane-wave solution, such zigzag ribbons provide a two-fold degenerate quasi-flat band of states localized to the edge.
Following the notation of Refs.~\cite{Amini1, Amini2} we can write the corresponding wave-functions and spectra for the edge states localized at the top edge of ribbon (which belongs to the $A$-sublattice) as
\begin{equation}\label{E5}
|\Psi^A(k) \rangle=\frac{1}{\sqrt{\pi}}\sum_{m,n}\gamma(k)\alpha^{n+\delta}(k)e^{ikm}|m,n,A \rangle, 
\end{equation}
and
\begin{equation}\label{E6}
E(k)= 4(2t_{4}t_{1})/t_{2} \cos^{2}(k/2)= 4t^{'}\cos^{2}(k/2)=2\acute{t}\cos k-2,  
\end{equation}
respectively
with $\gamma^{2}(k)=1-\alpha^{2}(k)$ and $\alpha(k)=-2\frac{t_{1}}{t_{2}}\cos(\frac{k}{2})$.  
Here $|m,n,\nu \rangle$ represent the atomic states localized on the Phosphorous atoms at position $(m,n,\nu)$ (see Fig.~\ref{F1} ) and
wave-vector $k$ measured in units of inverse lattice spacing $a$. Also, $\delta$ is a constant value equal to $0 (0.5)$ for even (odd) $n$.  

\subsection{Localized states around two nearby vacancies}

We are now in a position to focus on the localized states associated with vacancy defects. 
In what follows, vacancies can be considered by setting all the hopping
amplitudes to zero for the vacant sites.
Theoretical calculation~\cite{Amini2}, as well as the experimental observation~\cite{exp}, shows that 
when a single vacancy is created in bulk Phosphorene, a highly anisotropic wave-function appears to be localized spacially around it.
Without loss of generality, let us consider the case that a single vacancy is located on site $(n_{0},0, B)$ where $n_0\gg1$.
The exact analytical wave-function associated with such localized state induced by a single vacancy can be expressed as a combination of edge states as~\cite{Amini2} 
\begin{equation}\label{E7}
|\Psi_1(k) \rangle=c\int^{\pi}_{-\pi}\gamma^{-1}(k)|\Psi^A(k)\rangle dk,  
\end{equation}
with the following energy
\begin{equation}\label{E9}
E=c^{2}\int^{\pi}_{-\pi}\gamma^{-2}(k)4t^{'}\cos^{2}(k/2) dk,  
\end{equation}
where the normalization factor $c$ satisfies $c^{-2}=\int^{\pi}_{-\pi}\gamma^{-2}(k)dk$.

Now it is straightforward to generalize this formalism to the case where we have a set of two nearby vacancies in the bulk.
When we have two nearby vacancies, their corresponding localized states are coupled through the lattice.
This gives rise to the formation of a symmetric and an anti-symmetric localized state. 
To obtain their energy levels and wave-functions we consider a second vacancy to be located on site  $(n_{0},m_0,B)$.
Therefore, the new vacancy state reads exactly as
\begin{equation}\label{E10}
|\Psi_{2}\rangle =c\int^{\pi}_{-\pi}\gamma^{-1}(k)e^{ikm_{0}}|\Psi(k)\rangle dk. 
\end{equation}
with the same energy.
We should note, however, that such states are not orthogonal to one another since $\langle\Psi_{1}|\Psi_{2}\rangle\neq 0$.
Therefore, we can introduce a sym­met­ric and an ant-symmetric com­bi­na­tion of the two localized states  as
\begin{eqnarray}\label{E12}
|\Psi^{+}\rangle&=&c^{+}(|\Psi_{1}\rangle+|\Psi_{2}\rangle), \\ \nonumber
|\Psi^{-}\rangle&=&c^{-}(|\Psi_{1}\rangle-|\Psi_{2}\rangle),  
\end{eqnarray}
in which the  normalization coefficients $c^{\pm}$  are
\begin{equation}\label{E13}
(c^{\pm})^{2}=\frac{1}{2(1\pm c)}.
\end{equation}

Since $\langle\Psi^{+}|H_{1}|\Psi^{-}\rangle=0$, it is obvious that their corresponding energy levels can be written as 
$\langle\Psi^{\pm}|H_{1}|\Psi^{\pm}\rangle=E^{\pm}$   
which yields as
\begin{equation}\label{E15}
E^{\pm}=(c^{\pm})[2E \pm \frac{2c^2}{2\pi}\int^{\pi}_{-\pi}\gamma^{-2}(k)4t^{'}\cos^{2}(k/2)\cos(km_{0})dk],  
\end{equation}
where $E$ is expressed in Eq.(\ref{E9}).
Therefore, such a pair of localized states due to the nearby vacancy defects in bulk Phosphorene can be described as a TLS embedded in the Phosphorene sheet.

\begin{figure}[t!]
  \centering
  \includegraphics[scale=0.4]{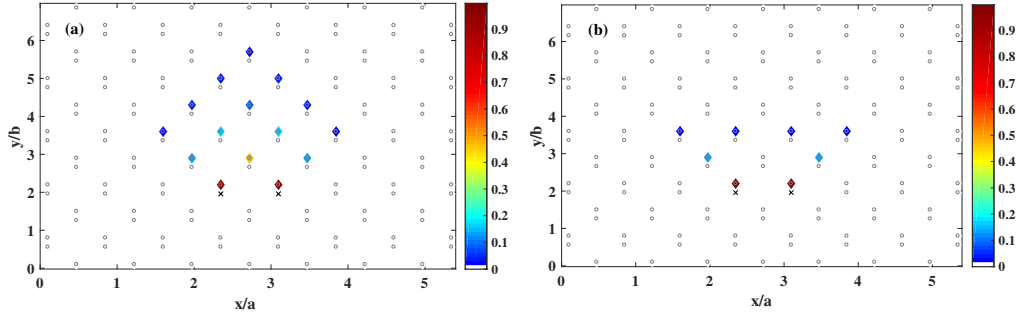}
  \caption{(a) shows the components of  wave-function  associated with the symmetric localized state around two nearby vacancies which are located on the sites denoted by $\times$ symbol in bulk Phosphorene and (b) shows the same thing for the anti-symmetric localized state. } \label{F2}
\end{figure}

Before ending this section, let us look at an example that shows such symmetric and anti-symmetric wave-function around two nearby vacancies in bulk Phosphorene. Fig.~\ref{F2} shows the amplitudes of wave-function on the lattice sites of Phosphorene in which the vacant sites are denoted by $\times$ symbol.

\section{Scattering of conduction electrons by a two-level system: generalized Lippmann-Schwinger formalism}\label{SecIII}
\subsection{General formulation}
In this sub-section, we introduce general formulation of electronic transport through a physical system with a single conducting channel 
which is side-coupled to a two-level quantum system formed by the presence of two impurity defects.
The minimal model for our purpose to study the basics of the Lippmann-Schwinger formalism
can be described by the following Hamiltonian
\begin{equation}\label{E16}
\hat{H}=\hat{H}_C+\hat{H}_{imp}+\hat{V}  
\end{equation}
where
\begin{equation}\label{E17}
\hat{H}_C=\int^{\pi}_{-\pi}E(k)|\Psi(k)\rangle \langle\Psi(k)|dk, 
\end{equation}
and
\begin{equation}\label{H_imp}
\hat{H}_{imp}=E_{1}|a_{1}\rangle \langle a_{1}| + E_{2}|a_{2}\rangle \langle a_{2}|.
\end{equation}
Here, $\hat{H}_C$ describes an effective one-dimensional continuum of extended states with energy $E(k)$ (with the Bloch wave number index $k$) and $\hat{H}_{imp}$ describes the localized impurity defects in which $|n\rangle$ and $E_n$ (with $n=1 ,2$)  correspond to two impurity states and their energy levels respectively.
$\hat{V}$ is responsible for the coupling between the localized states and the extended ones which takes the following general
form
\begin{equation}\label{E18}
\hat{V}=\int^{\pi}_{-\pi}(|\Psi(k)\rangle \langle \varphi(k)|+h.c.) dk, 
\end{equation}
where
\begin{equation}\label{E19}
|\varphi(k)\rangle =u_{1}(k)|a_{1}\rangle + u_{2}(k)|a_{2}\rangle.  
\end{equation}
In Eq.~(\ref{E19}) $u_l(k)$  is the spectral coupling
function between the continuum and the $l$-th discrete level.

Now, we can proceed to the scattering problem of an electron with energy $E$ in the continuum by the impurities.  
This can be described by the solution of Lippmann-Schwinger equation in which the scattering wave state $|\Psi\rangle$ obeys 
\begin{equation}\label{E20}
|\Psi\rangle = |\Psi_{0}\rangle + \hat{G_{0}}\hat{T}|\Psi_{0}\rangle,  
\end{equation}
where we can define the free Green’s function operator as
\begin{equation}\label{E22}
\hat{G_{0}}=\int^{\pi}_{-\pi}\frac{|\Psi(k)\rangle \langle \Psi(k)|}{E-E(k)+i0^{+}}dk + \frac{|a_{1}\rangle \langle a_{2}|}{E-E_{1}+i0^{+}}+\frac{|a_{2}\rangle \langle a_{2}|}{E-E_{2}+i0^{+}},  
\end{equation}
and the transition operator $\hat{T}$ as
\begin{equation}\label{E21}
\hat{T}(E)=\hat{V}+\hat{V}\hat{G_{0}}(E)\hat{V}+\hat{V}\hat{G_{0}}(E)\hat{V}\hat{G_{0}}(E)\hat{V}+\cdots. 
\end{equation}
  
Inserting Eqs.~(\ref{E21}) and (\ref{E22}) into Eq.~(\ref{E20}) and performing some algebra, it is straightforward
to see that the transition operator $\hat{T}$ takes the form 
\begin{equation}\label{E23}
\hat{T}=\hat{T_{1}}+\hat{T_{2}}  
\end{equation}
where the first term describes the transition of the electrons from the free Bloch stats $|\Psi(k)\rangle$ to the impurity states as well as the inverse transition and can be expressed as
\begin{equation}\label{E24}
\hat{T_{1}}=\int^{\pi}_{-\pi}(|\varphi^{'}(k)\rangle \langle \Psi(k)|+h.c.)dk,  
\end{equation}
in which
\begin{equation}\label{E25}
|\varphi^{'}(k)\rangle =u^{'}_{1}(k)|a_{1}\rangle + u^{'}_{2}(k)|a_{2}\rangle  
\end{equation}
with
\begin{equation}\label{E26}
\left[
  \begin{array}{c}
    u^{'}_{1}(k) \\
    u^{'}_{2}(k) \\
  \end{array}
\right]
 = (\frac{1}{1-\hat{g}})\left[
                          \begin{array}{c}
                            u_{1}(k) \\
                            u_{2}(k) \\
                          \end{array}
                        \right].
\end{equation}
The $\hat{g}$ operator in Eq.~(\ref{E26}) is a square matrix whose elements defined as
\begin{eqnarray}
g_{11}&=&\int_{-\pi}^{\pi}\frac{|u_{1}(k)|^{2}}{(E-E(k))(E-E_{1})}dk \nonumber \\ 
g_{22}&=&\int_{-\pi}^{\pi}\frac{|u_{2}(k)|^{2}}{(E-E(k))(E-E_{2})}dk  \nonumber \\
g_{12}&=&\int_{-\pi}^{\pi}\frac{u_{1}^{*}(k)u_{2}(k)}{(E-E(k))(E-E_{1})}dk \nonumber \\ 
g_{21}&=&\int_{-\pi}^{\pi}\frac{u_{2}^{*}(k)u_{1}(k)}{(E-E(k))(E-E_{2})}dk.  
\label{E27}
\end{eqnarray}
The second term of Eq.~(\ref{E23}), $\hat{T_{2}}$, describes the transition from a state in the continuum to another state in it or from an impurity state to another one and can be expressed as
\begin{equation}\label{E28}
\hat{T_{2}}=\int_{-\pi}^{\pi}F(k,k^{'})|\Psi(k) \rangle \langle \Psi (k^{'} )|dkdk^{'}+\int_{-\pi}^{\pi} \frac{1}{E-E(k)} |\varphi^{'}(k)\rangle \langle \varphi(k)| dk,	  
\end{equation}
where
\begin{equation}\label{E29}
F(k,k^{'})= \left[
              \begin{array}{cc}
                u^{*}_{1}(k) & u^{*}_{2}(k) \\
              \end{array}
            \right]
            \left[
              \begin{array}{cc}
                \frac{1}{E-E_{1}+i0^{+}} & 0 \\
                0 & \frac{1}{E-E_{2}+i0^{+}} \\
              \end{array}
            \right]
            \left[
              \begin{array}{c}
                u_{1}(k) \\
                u_{2}(k) \\
              \end{array}
            \right]. 
\end{equation}

Finally, by combining Eq.~(\ref{E20}) and the definition of
transmission function,  we can obtain the transmission amplitude $t$ of an incident electron with the state  $|\Psi_{in}\rangle = |\Psi(k_{0})\rangle$ and energy $E_0=E(k_0)$ through the channel as
\begin{equation}\label{E31}
t=1+G_{0} (E_{0} )F(k_{0},k_{0}).	  
\end{equation}  
Once the transmission amplitude is known, it is easy to obtain the transmission probability as $T=|t^2|$.
This is what we will refer to in the next sub-sections.

\subsection{Transmission in a model quantum wire side-coupled to two different impurities}

\begin{figure}[t!]
  \centering
  \includegraphics[width=8cm]{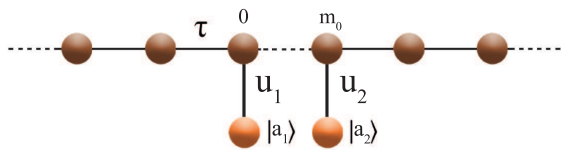}
  \caption{Schematic view of a one-dimensional
chain with hopping amplitude $\tau$ to the nearest neighbors. Two impurity states $ |a_{1} \rangle$ and $|a_{2} \rangle$ are coupled with amplitudes $u_{1}=\tau_1$ and $u_{2}=\tau_2$ to the zeroth and $m$-site of this chain respectively.}\label{F3}
\end{figure}

As an illustrative example of our model calculation, we first consider 
the simplest model which supports the coupling between two discrete states and a continuum of states. 
Such a model consists of a linear chain of identical sites (quantum wire) and two on-site impurities as shown schematically in Fig.~\ref{F3}. 
Using the standard tight-binding basis, the Hamiltonian of this system
can be written as
\begin{eqnarray}\label{E33}
\hat{H}&=&\hat{H}_C+\hat{H}_{imp}+\hat{V} \\ 
 &=&\tau \sum_{i} (c^{\dag}_{i}c_{i+1} + h.c.)  + E_{1}a^{\dag}_{1}a_{1}+E_{2}a^{\dag}_{2}a_{2} + \tau_1 (a^{\dag}_{1}c_{0} + h.c.) + \tau_2(a^{\dag}_{2}c_{m_{0}} + h.c.),  \nonumber
\end{eqnarray}
in which $c^\dagger(c)$ and $a^\dagger (a)$ represent the creation (annihilation) operators for the wire and the impurity respectively.
The distance between each lattice site of the chain and its nearest neighbor is considered to be unity and the electron-hopping amplitude between them is equal to $\tau$. 
At the zeroth ($m_0$-th) site of the chain an impurity state is attached with coupling $\tau_1 (\tau_2)$ and on-site energy $E_1 (E_2)$.

Due to the translational invariant of the one-dimensional chain, its corresponding term in  Eq.~(\ref{E33}) can be diagonalized by the
wave-number representation defined by
\begin{equation}\label{E36}
c_{i_{0}}^{\dag} |0 \rangle = \int_{-\pi}^\pi e^{iki_{0}} |\Psi(k)\rangle dk. 
\end{equation}
Therefore, according to the notation of Eq.~(\ref{E16}) and  with 
\begin{equation}\label{Energy}
E(k) = −2\tau \cos(k) 
\end{equation}
as dispersion relation of an electron in the continuum with wave-number $k$, we have
\begin{equation}\label{E34}
 \hat{H}_C+\hat{H}_{imp}=\int^{\pi}_{-\pi}E(k)|\Psi(k)\rangle \langle\Psi(k)|dk + E_{1}|a_{1}\rangle \langle a_{1}| + E_{2}|a_{2}\rangle \langle a_{2}| 
\end{equation}
and
\begin{equation}\label{E37}
\hat{V}=\int _{-\pi}^{\pi} (u_{1}(k)|\Psi(k)\rangle \langle a _{0}(k)| + u_{2}(k)|\Psi(k)\rangle \langle a _{m_{0}}(k)| + h.c.)dk  
\end{equation}
where
\begin{eqnarray}
\label{E38}
u_{1}(k) &=& \tau_1 \nonumber \\  
u_{2}(k) &=& e^{ikm_{0}}\tau_2.  
\end{eqnarray}
Substituting Eqs.~(\ref{E38}) and ~(\ref{Energy}) into Eq.~(\ref{E27}), we will obtain the following matrix elements
\begin{eqnarray}
g_{11}&=&\frac{1}{2i\tau_1\sin(k_{0})} \times \frac{1}{E-E_{1}}\nonumber \\
g_{22}&=&\frac{1}{2i\tau_2\sin(k_{0})} \times \frac{1}{E-E_{2}}\nonumber \\ 
g_{21}&=&\frac{e^{ik_{0}m_{0}}}{2i\sqrt{\tau_1\tau_2}\sin(k_{0})} \times \frac{1}{E-E_{1}} \nonumber \\
g_{12}&=&\frac{e^{-ik_{0}m_{0}}}{-2i\sqrt{\tau_1\tau_2}\sin(k_{0})} \times \frac{1}{E-E_{2}},  
\end{eqnarray}
from which we can evaluate Eqs.~(\ref{E28}) and (\ref{E29}).

\begin{figure}[h!]
  \centering
  \includegraphics[scale=0.55]{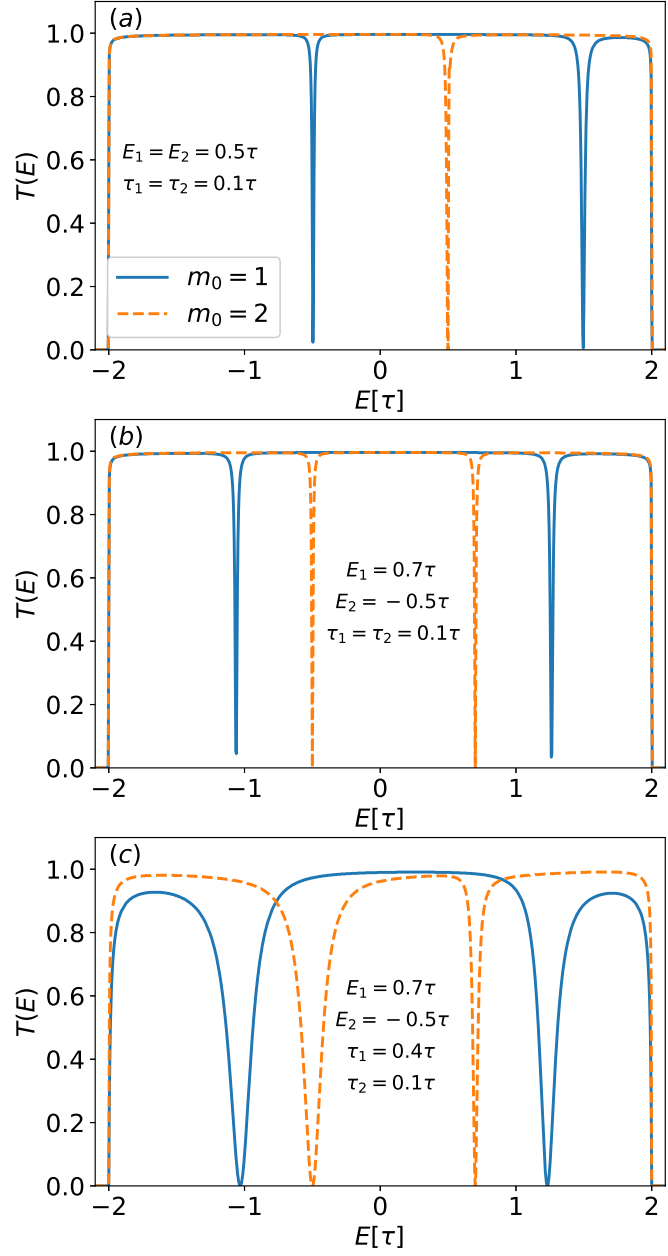}
  \caption{Transmission spectra $T(E)$ through a one-dimensional chain side-coupled to two different impurities which is shown in Fig.~\ref{F3} 
for $m_0=1,2$ when $\tau_1=\tau_2$ and $E_1=E_2$ (a), $\tau_1=\tau_2$ and $E_1\neq E_2$ (b),  and $\tau_1 \neq \tau_2$ and $E_1\neq E_2$ (c).}\label{F4}
\end{figure}


We now proceed to use Eq.~(\ref{E31}) to investigate the transmission features of this system analytically.
Let us consider that one of the impurities is coupled at the origin (zeroth site of the chain) whereas the other one is coupled to a site
located a distance $m_0$ away from the first impurity. 
To start with, we discuss the results for the two identical impurity case where $\tau_1=\tau_2=0.1\tau$ and $E_1=E_2=0.5 \tau$.
In Fig.\ref{F4} (a), we present some graphical representations of transmission probability $T$ as a
function of the energy when the two impurities are attached to two different sites of the chain 
located one ($m_0=1$) or two ($m_0=2$) sites away. 
As can be seen, when $m_0=2$ the transmission curve shows two narrow dip structures with a significant relative energy shift which can be viewed as Fano line shapes around the double-dip profile.
On the other hand, when $m_0=1$, the transmission spectra exhibit only a single dip forming a single Fano (anti-)resonance.
In Figs. \ref{F4}(b)–(c), we illustrate the same thing but obtained for different values
of the coupling strength $\tau_1$ and $\tau_2$ as well as the on-site potentials $E_1$ and $E_2$. 
It is seen that if we consider different on-site impurity energies $E_1$ and $E_2$ (e. g. $E_1=0.7\tau$, and $E_2=-0.5\tau$) we can have the double-dip profile for both situations $m_0=1$ and $m_0=2$ which is shown in Fig~\ref{F4} (b).
The most noticeable thing will takes place when the coupling strengths $\tau_1$ and $\tau_2$ are chosen
differently (e. g. $\tau_1=2\tau_2=0.4\tau$).
This is shown in Fig~\ref{F4} (c) in which we see that the dip structures become more broadened.
This results in the appearance of more asymmetric line shapes, which are typical for the Fano resonances.


\subsection{Transmission through the edge channel of ZPNRs in the presence of two nearby vacancies}
We next aim to study the Fano double-dip profile in the case of defective ZPNRs.
As we already discussed in Sec.~\ref{SecII},   
the creation of a set of two nearby vacancies leads to the formation of two new localized states around the vacancies in bulk Phosphorene.
It is now interesting to discuss the situation in which we create a pair of vacancies in a region that is close to the boundary of a ZPNR.
Now, the pair of localized states induced by the two vacancy defects can couple to the continuum of states formed by the edge states of the ZPNR. Therefore, we need to replace the Bloch wave states $|\Psi(k)\rangle$ and its spectrum in Eq.~(\ref{E17}) by the edge states of Eq.~(\ref{E5}) and their corresponding energy dispersion of Eq.~(\ref{E6}).
Furthermore, the impurity states $|1\rangle$ and $|2\rangle$ should be replaced by the sym­met­ric and anty-symmetric wave functions of the two-level system which is obtained in Eq.~(\ref{E12}).         
The only remaining thing is that the edge states of Eq.~(\ref{E5})  and the localized states of Eq.~(\ref{E12}) are not orthogonal, namely, $\langle \Psi(k) | \Psi^\pm(k) \rangle \neq 0$.
Therefore, we need to make new localized states which are orthogonal both to the edge states or to each other.
This can be done following~\cite{Amini2} by defining the following localized states
\begin{equation}\label{E41}
|\Psi^{'}_{i}\rangle=c^{'}_{i}(|\Psi_{i}\rangle-\int^{\pi}_{-\pi}\langle\Psi_{i}|\Psi(k)\rangle|\Psi(k)\rangle dk),  
\end{equation}
where $|\Psi_1\rangle$ and $|\Psi_2\rangle$ are given by Eqs.~(\ref{E7}) and~(\ref{E10}). 
Here, $c^{'}$ is the normalization coefficient which is written as
\begin{equation}\label{E43}
c^{'2}=1-\int^{\pi}_{-\pi}|\langle\Psi_{i}|\Psi(k)\rangle|^{2}dk.  
\end{equation}
It is now obvious that this wave function satisfies the following condition
\begin{equation}\label{E42}
\langle\Psi(k)|\Psi^{'}_{i}\rangle=0.  
\end{equation} 
Then we can easily make the new orthogonal symmetric and antisymmetric state as
\begin{equation}\label{E44}
|\Psi^{\pm}\rangle=c^{'\pm}(|\Psi^{'}_{1}\pm |\Psi^{'}_{2}\rangle) 
\end{equation}
with the following normalization factor
\begin{equation}\label{E45}
c^{'2}_{\pm}=2c(1\pm \int^{\pi}_{-\pi}|\langle\Psi_{i}|\Psi(k)\rangle^{2}\cos(km_{0})dk).  
\end{equation}
Since $\langle\Psi^{+}|H_{1}|\Psi^{-}\rangle=0 $ and $\langle \Psi ^{\pm}|H_{1}|\Psi^\pm \rangle=E^{\pm}$, 
one can obtain the the corrsponding bonding energies of such localized states which are given by  
\begin{equation}\label{E48}
E^{\pm}=c^{'2}_{\pm}(E\pm E^{'})  
\end{equation}
where  
\begin{eqnarray}\label{E49}
E&=&E_{00}+2E_{01}+E_{11} \\ \nonumber
E^{'}&=&E^{'}_{00}+2E^{'}_{01}+E^{'}_{11}.   
\end{eqnarray}
If we consider the vacancies to be located on the $n_0$-th zigzag chain of the ribbon, we have
\begin{eqnarray}
E_{00}&=&c^{2}\int^{\pi}_{-\pi}\gamma^{-2}(k)4t^{'}\cos^{2}(k/2)dk \\ \nonumber
E_{01}&=&c^{2}\int^{\pi}_{-\pi}\gamma^{-2}(k)4t^{'}\cos^{2}(k/2)[\alpha^{2n_{0}-2}(k)+\alpha^{2n_{0}}] dk \\ \nonumber
E_{11}&=&c^{2}\int^{\pi}_{-\pi}\gamma^{-2}(k)4t^{'}\cos^{2}(k/2) \alpha^{2n_{0}} dk,     
\end{eqnarray}
and
\begin{equation}
E^{'}_{ij}=E_{ij}\cos(km_{0}).
\end{equation}


To close the problem, we need only to find the coupling functions $u_i(k)$ which can be calculated using  $u_{1(2)}(k) = \langle \Psi (k) | H_{1} | \Psi^{+(-)} \rangle$ which yields
\begin{eqnarray}
u_{1}(k)&=&2c^{'}_{+}e^{ikm_{0}/2}u(k)\cos (m_{0}/2), \nonumber  \\  
u_{2}(k)&=&2c^{'}_{-}e^{ikm_{0}/2}u(k)\sin (m_{0}/2)  
\end{eqnarray}
in which
\begin{equation}\label{E54}
u(k)=c^{'2}\gamma^{-2}(k)\frac{t_{2}t_{4}}{t_{1}}[\alpha^{2n_{0}}(k)-2\alpha^{2n_{0}+2}(k)+\alpha^{2n_{0}+4(k)}].  
\end{equation}
Similar to the analysis described in the previous sub-section, these equations used to calculate the matrix elements of $\hat{g}$ in Eq.~(\ref{E27})  to find the value of transmission amplitude in Eq.~(\ref{E31}).
In what follows we present some graphical representations of the transmission probability $T=|t|^2$ as a function of energy $E$ for 
different locations of double vacancies. 

\begin{figure}[t!]
  \centering
  \includegraphics[scale=0.53]{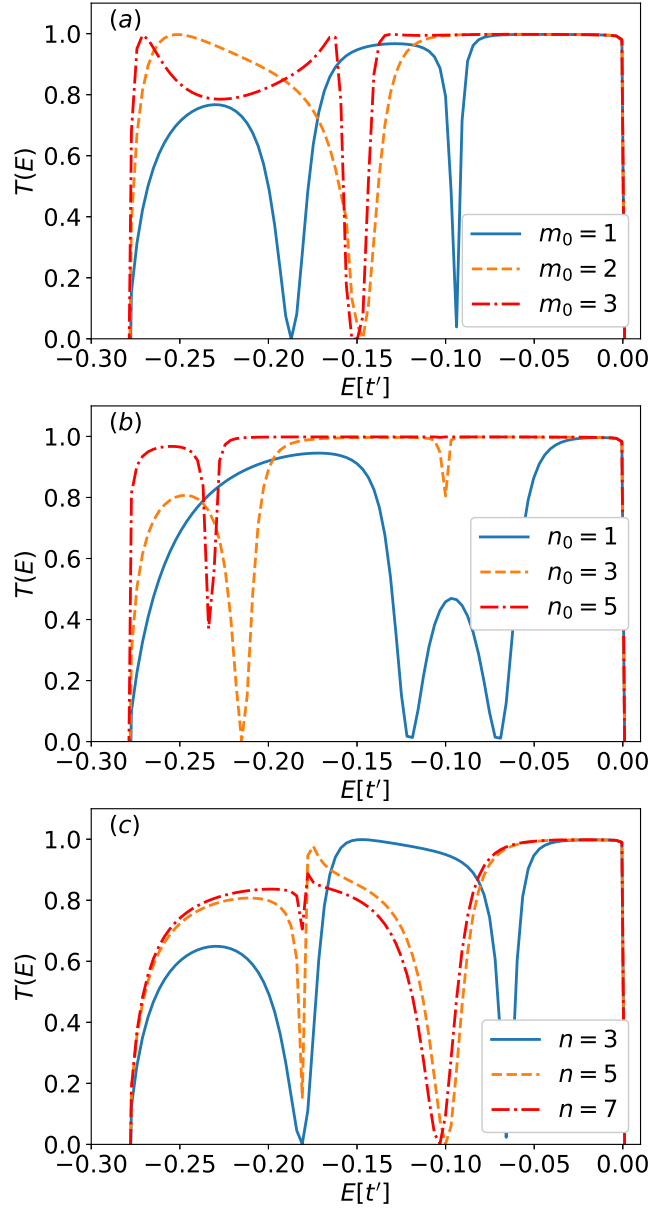}
  \caption{Transmission spectra of the proposed TLS embedded in a ZPNR  as a function energy for different positions of the vacancy impurities.  (a) shows the corresponding transmission for the case where two nearby vacancies are located on the same zigzag chain and have different intervacancy distance.  (b) shows the same thing for fixed intervacancy distance and a different distance from the edge. (c) corresponds to the case where vacancies are coupled in the vertical direction.}\label{F5}
\end{figure}

We initially restrict ourselves to the case where both vacancies are located on the same zigzag edge (belong
to the same sub-lattice) at the sites $(n_0=2,0,B)$ and $(2,m_0,B)$.
The resulting transmission probability $T(E)$ which is presented in Fig.~\ref{F5}-(a)
shows a double-dip Fano antiresonance when vacancies are very close to each other ($m_0=1$).  This is due to the formation of the symmetric and anti-symmetric bound states scenario which originates from the strong coupling of both vacancies. 
If we increase the distance between vacancies ($m_0=2,3$), the energy shift between the dips decreases, and one of the dips will disappear because this will decrease their coupling as expected. 
This is due to the existence of smaller energy splitting between symmetric and anti-symmetric states which takes place for the larger intervacancy distances.
Besides the distances between vacancies, it is also interesting to note how the Fano profile changes with changing the proximity to the edge of ZPNR. To see what happens now, we fix the intervacancy distance ($m_0=1$) and let the distance of the vacancies from the edge be $n_0=1,2,3$. The resulting transmission spectra are plotted in Fig.~\ref{F5}-(b).
As it is obvious, now there are two close dips when the vacancies are located near the edge ($n_0=1$) and by increasing their distance from the edge, the energy shift between the dips increases. Furthermore, one of the antiresonances became narrower and eventually disappear for $n_0=5$.

It is also possible to discuss the case where vacancies are side-coupled in the vertical direction. Let us consider the first vacancy at site $(1,0,B)$ and the the second one at site $(n,0,B)$. 
Figure ~\ref{F5}-(c) illustrates how the transmission spectra change when we change the position of the second vacancy along the vertical axis for $n=3, 5, 7$. In this case, the position of the dip at lower energy does not change more or less while the dip at higher energy moves to a lower energy position.

\section{Conclusion}\label{SecIV}

In summary, we have investigated electron scattering
properties in a quasi one-dimensional edge channel of the ZPNR in presence of a pair of vacancy defects analytically.
We found that the localized states associated with vacancies in the bulk Phosphorene form symmetric
and antisymmetric states when the vacancies are close to each other.
The energy splitting between these states decreases with increasing intervacancy distance as we expected.
This somehow offers a practical realization of a two-level quantum device in Phosphorene.
As an application for such a device, we have developed a theoretical model by implementing the Lippmann-Schwinger formalism to study the quantum transport properties of a such double-level system which is side coupled to the edge channel of a ZPNR.
Using this theoretical formalism, we calculated the transmission spectra for different positions of the two vacancies near the edge of the ZPNR.

\section*{References}

\begin{thebibliography}{99}
\bibitem{Fano} U. Fano, Phys. Rev. 124, 1866 (1961).
\bibitem{Waligorski} G. Waligorski, L. Zhou,  W. E. Cooke,  Phys. Rev. A 55, 1544–1547 (1997).
\bibitem{Limonov} M. F. Limonov, M. V. Rybin, A. N. Poddubny, and Y. S. Kivshar, Nat. Photonics 11, 543–554 (2017).
\bibitem{Bulka} B. R. Bulka and P. Stefanski, Phys. Rev. Lett. 86, 5128 (2001).
\bibitem{Lukyanchuk} B. Lukyanchuk,  et al.,  Nat. Mater. 9, 707–715 (2010).
\bibitem{Samson} Z. L. Samson, et al., Appl. Phys. Lett. 96, 143105 (2010).
\bibitem{Kobayashi-2004} K. Kobayashi, H. Aikawa, A. Sano, S. Katsumoto, and Y. Iye, Phys. Rev. B 70, 035319 (2004).
\bibitem{Kobayashi-2002} K. Kobayashi, H. Aikawa, S. Katsumoto, and Y. Iye, Phys. Rev. Lett. 88, 256806 (2002).
\bibitem{Kobayashi-2005} M. Sato, H. Aikawa, K. Kobayashi, S. Katsumoto, and Y. Iye, Phys. Rev. Lett. 95, 066801 (2005).
\bibitem{Johnson} A. C. Johnson, C. M. Marcus, M. P. Hanson, and A. C. Gossard, Phys. Rev. Lett. 93, 106803 (2004).
\bibitem{Orellana} P. A. Orellana, G. A. Lara, and E. V. Anda, Phys. Rev. B 74, 193315 (2006).
\bibitem{Ricco} L. S. Ricco, V. L. Campo Jr., I. A. Shelykh, and A. C. Seridonio, Phys. Rev. B 98, 075142 (2018).
\bibitem{Yang} P. Y. Yang and W. M. Zhang, Phys. Rev. B 97, 054301 (2018).
\bibitem{Ladron} M. L. Ladron de Guevara and P. A. Orellana, Phys. Rev. B 73, 205303 (2006).
\bibitem{Wigner} J. von Neumann and E. Wigner, Phys. Z. 30, 465 (1929).
\bibitem{Capasso} F. Capasso, C. Sirtori, J. Faist, D. L. Sivco, S. G. Chu,  and A. Y. Cho,  Nature 358, 565 (1992).
\bibitem{Claro} M. L. Ladron de Guevara, F. Claro, and P.A. Orellana, Phys. Rev. B 67, 195335 (2003).
\bibitem{Sablikov} V. A. Sablikov and A. A. Sukhanov, Phys. Lett. A 379, 1775 (2015).
\bibitem{Sadreev} A. F. Sadreev, E. N. Bulgakov, and I. Rotter, Phys. Rev. B 73,
235342 (2006).
\bibitem{Albo} A. Albo, D. Fekete, and G. Bahir, Phys. Rev. B 85, 115307
(2012).
\bibitem{Longhi} S. Longhi and G. D. Valle, Sci. Rep. 3, 2219 (2013).
\bibitem{Tanaka} S. Tanaka, S. Garmon, G. Ordonez,  and T. Petrosky,  Phys. Rev. B 76, 153308 (2007).
\bibitem{Garmon} S. Garmon, H. Nakamura, N. Hatano, T. Petrosky - Phys. Rev. B, 80, 115318 (2009).
\bibitem{Reyes} S. A. Reyes, D. Thuberg, D. Pérez, C. Dauer, and S. Eggert, New J. Phys. 19, 043029 (2017).  
\bibitem{Bahamon} D. A. Bahamon, A. L. C. Pereira,  and P. A. Schulz, Phys. Rev. B 82, 165438 (2010).
\bibitem{Ezawa} M. Ezawa, New J. Phys. 16, 115004 (2014).
\bibitem{Asgari} Z. Nourbakhsh and R. Asgari, Phys. Rev. B 97, 235406 (2018).
\bibitem{Peeters} L. L. Li and F. M. Peeters, Phys. Rev. B 97, 075414 (2018).
\bibitem{Amini1} M Amini, M Soltani, Journal of Physics: Condensed Matter 31 (21), 215301 (2019).
\bibitem{Amini2} M Amini, M Soltani, M Sharbafiun, Phys. Rev. B 99 (8), 085403 (2019).
\bibitem{Katsnelson} A. N. Rudenko and M. I. Katsnelson, Phys. Rev. B 89, 201408(R) (2014).
\bibitem{exp} B. Kiraly, N. Hauptmann, A. N. Rudenko, M. I. Katsnelson, and A. A. Khajetoorians, Nano Lett. 17, 3607 (2017).
\end{thebibliography}

\end{document}